\documentclass[%
 final,
]{an}
\usepackage{graphicx}
\usepackage{times}
\begin{document}

\Pagespan{1}{}
\Yearpublication{2013}%
\Yearsubmission{2012}%
\Month{10}%
\Volume{334}%
\Issue{1-2}%
 \DOI{This.is/not.aDOI}%

\title{M dwarf stars in the light of (future) exoplanet searches}

\author{B. Rojas-Ayala\inst{1}\fnmsep\thanks{Corresponding author:
  \email{babs@amnh.org}}
\and  E.\,J. Hilton\inst{2,3}
\and A.\,W. Mann\inst{2}
\and S. L\'epine\inst{1}
\and E. Gaidos\inst{3}
\and X. Bonfils\inst{4}
\and Ch. Helling\inst{5}
\and T.\,J. Henry\inst{6}
\and  L.\,A. Rogers\inst{7}
\and K. von Braun\inst{8}
\and A. Youdin\inst{9}
}
\titlerunning{M dwarfs and exoplanets}
\authorrunning{B. Rojas-Ayala et al.}
\institute{
Department of Astrophysics, American Museum of Natural History, Central Park West at 79th Street, New York, NY 10024,
USA
\and Institute for Astronomy, University of Hawaii, Honolulu, HI 96822, USA
\and Department of Geology and Geophysics, University of Hawaii, Honolulu, HI 96822, USA 
\and UJF-Grenoble 1/CNRS-INSU, Institut de PlanŽtologie et d'Astrophysique de Grenoble (IPAG) UMR 5274, 38041, Grenoble, France
\and SUPA, School of Physics \& Astronomy, University of St. Andrews, North Haugh, St Andrews Ky16 9SS, UK
\and Department of Physics and Astronomy, Georgia State University, Atlanta, GA 30302-4106, USA
\and Department of Physics, Massachusetts Institute of Technology, Cambridge, MA 02139, USA
\and NASA Exoplanet Science Institute, California Institute of Technology, MC 100-22, Pasadena, CA 91125, USA
\and Harvard-Smithsonian Center for Astrophysics, 60 Garden St., Cambridge, MA 02138, USA  }

\received{XXXX}
\accepted{XXXX}
\publonline{XXXX}


\abstract{%
We present a brief overview of a splinter session on M dwarf stars as planet hosts that was organized as part of the Cool Stars 17 conference. The session was devoted to reviewing our current knowledge of M dwarf stars and exoplanets in order to prepare for current and future exoplanet searches focusing in low mass stars. We review the observational and theoretical challenges to characterize M dwarf stars and the importance of accurate fundamental parameters for the proper characterization of their exoplanets and our understanding on planet formation.}

\maketitle

\section{Motivation}
M dwarf stars have become a {\it hot} topic in the field of {\it cool stars}, in large part because of the interest in discovering small, rocky planets around them: smaller planets can be detected around stars with smaller radii (by the transit technique) or lower mass (by the Doppler technique). Planets can be very close to low luminosity M dwarfs but still be within the zone where an Earth-like planet could have liquid water. However, there is much we need to learn about M dwarfs, including the location of most of the nearest ones, the precise relationships between mass and radius,  and how to determine their metallicities and effective temperatures. Planet hunters need to know where to look, and how to scale the size of the surveys; determinations of the radius of a transiting planet are limited by the precision with which we know the radius of the host star; tests of planet formation theories involving metallicity are limited by how well we can measure metallicity.

\section{M dwarf Targets for Exoplanet Searches}

The search for exoplanets aims to improve on our knowledge of how do planets form, what are their structures and compositions, and, with perhaps a grander philosophical reach, what is the origin of life. We want exoplanet searches to detect as many (and as diverse) planets as possible and to detect the ones that can support the emergence of life. Today, the detection of habitable Earth-like planets transiting M dwarfs is one of the most appealing objectives since, aided by transmission and occultation spectroscopy, it is contemplated to be the shortest route to peek into an exo-life laboratory.

 The stellar members of the solar neighborhood are dominated by the red dwarfs, which comprise at least 74$\%$ of all stars within 10 pc (Henry
et al. 2006; www.recons.org). The RECONS group has been thoroughly canvassing the solar neighborhood in an effort to discover and characterize the Sun's nearest neighbors. To date, 131 new stellar systems have been published within 25 pc and another 188 are in the queue, most of them M dwarf stars. However, due to their intrinsic faintness, only a very small group of these nearby M dwarfs have been searched for exoplanets by magnitude limited ground-based surveys.

\begin{figure}[tb]
\includegraphics[width=\linewidth]{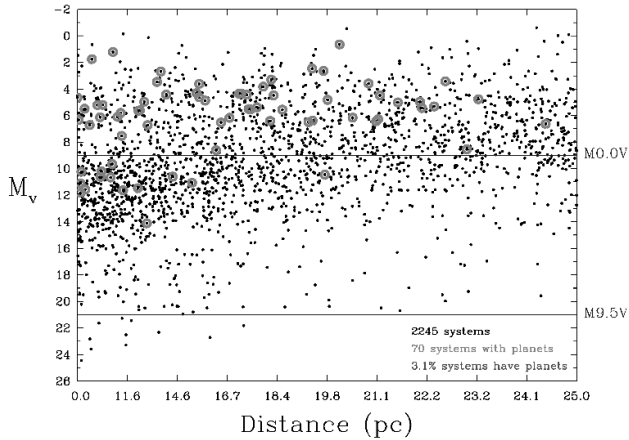}  
\caption{The 25 pc RECONS sample. The horizontal axis is designed to mark ten equal-volume shells moving away from the Sun, so that a constant density will appear as a constant number of points/bin. While the stellar sample appears to be complete for the closest AFGK stars, M dwarf stars are significantly underrepresented.}
\label{fig:todd}
\end{figure}

Shown in Fig. \ref{fig:todd} is a plot outlining what we know, and do not know, about the stars and exoplanets within 25 pc.  The 25 pc horizon is a convenient distance to consider because it should include a robust 6000 stellar systems,  yet only about 2000 have been confirmed with trigonometric parallax. Note that the stellar sample appears complete for AFGK stars brighter than M$_V$ = 9 mag.  Clearly, the M dwarfs with M$_V$ = 9 mag to 21 mag are significantly underrepresented. However, RECONS has made some progress at the closest distances, as evidenced by the accumulation of M dwarfs out to 15 pc.  Shown with gray circles are the 3.1$\%$ of stellar systems hosting exoplanet candidates, where the 14 planet hosts with M$_V$$>$9.0 mag correspond to early M dwarf stars. Mid to late M dwarfs in the solar neighborhood have barely been searched, due to their intrinsic faintness. In addition, surprisingly few exoplanets have been found around K dwarfs with M$_V$ = 6 mag to 9 mag.

In the past 10 years, surveys employing different techniques have been searching for planets around bright nearby M dwarfs as well as distant M dwarfs, and have recently reported on the occurrence of low-mass planets around M dwarfs. The HARPS M-dwarf survey have reported an occurrence rate of $88^{+56}_{-19}$\% for super-Earths ($1 \le m \sin i \le 10 {\rm M_\oplus}$) with periods $<100$ days, and $41^{+54}_{-13}$\% for those that orbit in the habitable zone (HZ; Bonfils et al. 2011). The Kepler mission also indicates a high occurrence rate for super-Earths with periods $<50$ days orbiting late-K to early-M dwarfs (3600$<T_{\rm eff}<4100$ K) with Howard et al. (2012) and Mann et al. (2012a) reporting an occurrence rate of $30\pm8$\% and $36\pm8$\%, respectively. The steep rise in the planetary mass function toward lower mass planets is given further credit by micro-lensing surveys which, for instance, find a 62$^{+35}_{-37}$\% occurrence rate for 5-10 ${\rm M_\oplus}$ planets at large separation (0.5-10 AU; Cassan et al. 2011). Last but not least, 2 posters at CS17 were also reporting on the occurrence rate of planets orbiting M dwarfs: Dressing \& Charbonneau (2012) for the occurrence of habitable planets from Kepler (0.40 habitable planet per star) and Berta et al. (2012) for the occurrence of 2-4 ${\rm R_\oplus}$ planets, with periods $P<10$ day (22$^{+52}_{-6}$\%) in the MEarth sample.

Hence, a consistent picture emerges from HARPS, Kepler, MEarth and micro-lensing surveys : the occurrence rate of $1-10 {\rm M_\oplus}$ planets is 30-50\% per d$\log P$, and there are 40 habitable-zone super-Earths for every 100 M dwarfs. One can just propagates these numbers to refine (or revise) the statistics and find that:
\begin{itemize}
\item the minimum sample size starts at N=15. With 15 stars, only a null result would be different than any one of the above surveys (with a 3-sigma confidence level). With 20 stars, one can refine the, e.g., HARPS determination of $\eta_{\oplus}$ by a factor of $\sim$2.
\item the minimum sample size to {\it expect} 1 transiting habitable super-Earth is $\sim$100 (and few times 100 to {\it confidently} detect one).
\end{itemize}

\section{The importance of M dwarf fundamental parameters for exoplanet searches}

To characterize a transiting exoplanet, one must first characterize its host star. Planet radius, mass, and $T_{\rm eq}$ are measured relative to the stellar properties. Uncertainties in the host star mass and radius often dominate the uncertainties on the planet mass and radius. A wide range of bulk compositions (with different proportions of rock, H/He, and ices) is consistent with the same planet mass-radius pair (e.g., Valencia, Sasselov \& O'Connell 2007b; Selsis et al. 2007; Rogers \& Seager 2010a), and therefore uncertainties in the planet mass and radius further broaden the span of possible compositions. Due to the inherent and observational uncertainties, the nature of a transiting low-mass planet is often ambiguous, with multiple plausible scenarios. For example, based on its measured mass and radius, GJ 1214b could be a mini-Neptune with a primordial H/He envelope, a water-dominated planet, or a rocky planet with an outgassed H-rich envelope (Rogers \& Seager 2010b). 

Improved precision on host-star properties will most significantly impact planet characterization in cases where reduced error bars on the planet properties could rule out a planet composition scenario. For instance, a planet with radius in excess of a pure silicate body of the same mass must contain significant quantities of volatiles (H/He or ices) -Ð a rock-dominated composition scenario is precluded. Other limiting mass-radius relations include radius lower limits for planets based on an extreme pure-iron composition (e.g. Seager et al. 2007) and collisional stripping-induced iron enhancement (Marcus et al. 2010),  and radius upper limits for rocky planets with outgassed envelopes (Rogers et al. 2011), water-planets without H/He envelopes, and planets with liquid water oceans below H-rich gas envelopes (Rogers 2012). Planets with measured properties spanning these limiting-composition mass-radius relations could offer the highest science return (in the form of improved composition constraints and stronger tests of planet formation) from the effort invested in shrinking $M_p$ and $R_p$ uncertainties with refined stellar parameters.

\subsection{Inferring M dwarf properties from models}

M dwarfs span a range of stellar parameters that includes youthful brown dwarfs and elderly low-mass stars: $T_{\rm eff}$= 2500-3900K and $\log g$=3.0-5.0. Their metallicity can be solar-like but also very low owing to the long life time of M dwarfs. To infer M dwarf properties from models will further be complicated by their large magnetic field strength of $10^3$ times that of the Sun. The determination of fundamental parameter like $T_{\rm eff}$, $\log g$, element abundances, mass, and  radii (or luminosity) seems most difficult for young M dwarfs because of the potentially still ongoing accretion, but also older M dwarfs are challenging, as magnetic fields play a role that we are only starting to appreciate.  

Different parameters are derived from observations by applying different models: synthetic spectra from atmosphere models (ATLAS: Castelli \& Kurucz 2004; MARCS: Gustafsson et al. 2008; PHOENIX: Hauschildt, Baron \& Allard 1997, Allard et al. 2001,  Dehn 2007, Helling et al. 2008, Witte et al. 2009, Rice et al. 2010;  Tsuji 2005; Ackerman \& Marley 2001, Saumon \& Marley 2008; Burrows et al. 2001) are used to infer $T_{\rm eff}$, $\log g$ and element abundances. The masses, radii, and ages are inferred from comparison to evolutionary models (see Table 1 in Southworth 2009; Dotter et al. 2008; Chabrier et al. 2000; Burrows et al. 2001; Ventura et al. 1998).

A good strategy for fitting observational data with models is to use as many models as one can. Examples are presented in Dupuy et al. (2010) and Patience et al. (2012) who furthermore demonstrate the uncertainties resulting from fitting the whole observed spectral range vs. fitting individual observed spectral intervals. Despite all wishful thinking, the uncertainties in $T_{\rm eff}$ easily amount to 200K amongst the different atmosphere models and for $\log g$ to one order of magnitude. This uncertainty applies also if observational 'holes' are filled in with synthetic spectra as pointed out by Emily Rice during this Splinter session. The detailed analysis of the element abundances in M dwarf atmosphere is challenged by the start of dust formation which impacts already at $T_{\rm eff}$ as high as 2800K (Witte et al. 2009). A detailed assessment of uncertainties between the different atmosphere model families is not available to date, less so for evolutionary models which carry the atmosphere as outer boundary.  Only Sinclair et al. (2010) and Plez (2011) incorporate comparisons between model results that are interesting for M dwarfs.

\subsection{Direct measurements of M dwarf radii}

Direct determinations of M dwarf radii are primarily performed via studies of eclipsing binaries (EBs) or via measurements using long-baseline, optical/near-IR interferometry. The recent survey by Boyajian et al. (2012) doubles the number of high-precision, interferometrically measured M dwarf radii in the literature and constitutes further evidence for the ongoing discrepancy between model and empirical M dwarf radii. Boyajian et al. (2012) finds very weak to non-existent correlation with metallicity for bilateral relations between M dwarf radii, temperatures, and luminosities, in contrast to a clear metallicity dependence for equivalent relations involving broadband color indices. Boyajian et al. (2012) conclude that there is no systematic discrepancy for a fixed mass between directly determined single M dwarfs and their EB component counterparts. Knowledge of stellar radius, effective temperature, and luminosity, along with literature time-series, can provide a largely model-independent astrophysical and orbital characterization of the star-planet system for transiting planets (von Braun et al. 2012). Additionally, the location and extent of the system HZ, based on the measured astrophysical stellar parameters, determines whether any of the exoplanets (or perhaps their moons), could host liquid water on their surfaces (von Braun et al. 2011).

\section{Planet formation theory and metallicity}

Metallicity is central to modern theories of planet formation because heavy elements in the disk make dust (Youdin \& Kenyon 2012). To first approximation, stellar metallicity should reflect the disk's initial metallicity. However the disk's metallicity -- defined as a surface density ratio of solid particles to gas -- changes as particles and gas evolve separately (Youdin 2010).

The metallicity dependence of planetesimal formation is particularly important as a first step that depends on disk gas (Chiang \& Youdin 2010).  Models of particle sedimentation to the mid-plane first showed that solar metallicity could be a threshold for planetesimal formation (Sekiya 1998; Youdin \& Shu 2002).  This seemingly fortuitous coincidence depends on the gas disk's radial pressure gradients, which affect the vertical stirring and radial drift of solids, especially below meter-sizes.  This radial drift gives rise to ``streaming instabilities" (SI) in protoplanetary disks (Youdin \& Goodman 2005). Simulations show that SI can aerodynamically concentrate solid particles (Youdin \& Johansen 2007; Johansen \& Youdin 2007).  With sufficient grain growth,  solar metallicity is the threshold for strong particle clumping by SI and subsequent gravitational collapse into planetesimals (Johansen,Youdin \& Mac Low 2009).  Other mechanisms for triggering planetesimal formation in gas rich disks also show a strong (positive) correlation with disk metallicity (Youdin 2011; Shariff \& Cuzzi 2011).
The early phases of planet formation can help explain observed patterns of exoplanets versus stellar mass and metallicity.  A high metallicity and/or mass star with (presumably) corresponding disk properties favors early formation of planetesimals and the subsequent growth of  gas giants by core accretion.   Around low metallicity stars, planetesimal formation requires time to enrich the disk's metallicity, favoring super-Earths over giants.  For low mass stars, disk mass may be too low to form gas giants very often and the metallicity advantage may primarily benefit super-Earths.  

M-dwarf optical spectra are dominated by highly structured molecular absorption and standard abundance analyses based on high S/N and resolution optical spectra are only possible for very early M dwarfs and without the accuracy achieved in FGK stars (e.g. Bean et al. 2006) . In the past decade, several alternative techniques have been developed to estimate the metallicity of M dwarfs. There are photometric techniques based on the $V-K_s$ color metallicity dependence (e.g. Neves et al. 2012 and references therein), as well as spectroscopic techniques based on broad molecular features (Woolf \& Wallerstein 2006), NIR absorption features in modest resolution spectra (Rojas-Ayala et al. 2012; Terrien et al. 2012), and high-res $J$-band spectra ({\"O}nehag et al. 2012). While these techniques have limitations and most of them rely on FGK+M binary systems, they assign consistent metallicity estimates to the nearby M dwarf planet hosts ($\sigma$[Fe/H]$ \sim$0.1-0.2 dex). The [Fe/H] estimates for the M dwarf planet hosts are consistent with the [Fe/H] distribution of FGK planet hosts, where jovian planets are preferentially found around metal-rich stars (e.g Rojas-Ayala et al. 2012, Muirhead et al. 2012). Current M dwarf exoplanet surveys are using these and new techniques to determinate the metallicities of several nearby M dwarfs (e.g. Newton et al. 2012, Mann et al. 2012b, Montes et al. 2012, Neves et al. 2012) which are essential to discriminate among theoretical planet formation scenarios.


\acknowledgements
The authors thank the CS17 SOC for the opportunity to hold this productive splinter session and the anonymous referee for her/his constructive comments. ChH highlights financial support of the European Community under the FP7 by an ERC starting grant. BR gratefully acknowledges the CS17 Early Career Travel Grant provided by the NASA Astrobiology Institute.

\end{document}